# Structural, dielectric and magnetic studies of (0–3) type multiferroic (1-x) BaTi$_{0.8}$Sn$_{0.2}$O$_3$–(x) La$_{0.5}$Ca$_{0.5}$MnO$_3$ (0<x<1) composite ceramics

S. Ben Moumen[1*], Y. Hadouch[1], Y. Gagou[2], D. Mezzane[1], M. Amjoud[1], E. Choukri[1], Z. Kutnjak[3], B. Rožič[3], N. Abdelmoula[4], H. Khemakhem[4], Y. El Amraoui[5, 6], V. Laguta[7, 8] and M. El Marssi[2].

[1]*IMED-Lab, F.S.T.G. Cadi Ayyad University, BP 549, Marrakech, Morocco*
[2]*LPMC, University of Picardie Jules Verne, 33 rue Saint-Leu, 80039 Amiens Cédex, France*
[3]*Jozef Stefan Institute, Jamova cesta 39, 1000 Ljubljana, Slovenia*
[4]*Laboratory of Multifunctional Materials and Applications (LaMMA), (LR16ES18), Faculty of Sciences of Sfax, University of Sfax, B. P. 1171, 3000 Sfax, Tunisia*
[5]*LaMCScI, Faculty of Sciences, Mohammed V University, Rabat, Morocco*
[6] *National School of Arts and Crafts, Moulay Ismail University, Meknes, Morocco.*
[7]*Institute of Physics AS CR, Cukrovarnicka 10, 162 53 Prague, Czech Republic*
[8]*Institute for Problems of Materials Science, National Ac. of Science, Krjijanovskogo 3, 03142 Kyiv, Ukraine*

*Corresponding author's email: said.benmoumen@yahoo.fr

## Abstract

Multiferroic composites with the composition (1-x) BaTi$_{0.8}$Sn$_{0.2}$O$_3$ - (x) La$_{0.5}$Ca$_{0.5}$MnO$_3$ (x =0, 0.3, 0.5, 0.7, 0.9 and 1) were prepared by sol-gel techniques. Rietveld refinements of the composites with 0-3 and 3-0 type connectivities exhibit the formation of pure phases and confirm the fraction of the composite samples. It was found that the composites grain size decreases slightly with increasing the La$_{0.5}$Ca$_{0.5}$MnO$_3$ content. Magnetic measurements of the composite ceramics exhibit a gradual increase in saturation and remanent magnetization with increasing La$_{0.5}$Ca$_{0.5}$MnO$_3$ content. Dielectric data demonstrate that the composite materials show high values of the dielectric constant in comparison with the parent compound BaTi$_{0.8}$Sn$_{0.2}$O$_3$ (BTSO). In addition, an attempt is made to measure the magnetoelectric coupling coefficient and to explain the negative permittivity and losses occurring in these materials at a certain frequency and temperature conditions.

## 1. Introduction

The continuous demand for miniaturization of electronic components stimulates researchers to identify and develop new materials that exhibit multifunctional properties at the nanoscale level. Multiferroic materials exhibiting both ferromagnetism and ferroelectricity or ferroelasticity, have attracted much attention due to the magnetoelectric effect in which the polarization/magnetization can be changed by applying an external magnetic/electric field, respectively. This phenomenon is leading to a wide range of applications such as multiple-stage memories, magnetoelectric transducers, sensors and actuators [1–3]. So far, single-phase multiferroics are of limited number [4–6]. So other methods have been followed for the development of new heterojunction materials [7], laminated composites [8,9], or composite materials [10,11] using the soft chemistry to achieve the nanometric scale of the grains in a way to enhance the mechanical coupling between the various elements of the composite compared with materials presenting coarser grains [12].

Moreover, the decrease in the grain size can generate strong ferromagnetic interaction between the grains at the nanoscale level and, as a result, the temperature transition can be affected [13,14] and the new magnetic orders can be observed [15]. Therefore, researchers are lately focused on the production of nano ferromagnetic and ferroelectric phases in which the combination leads to composites with excellent magnetic and dielectric performances [16]. Sol-gel technique has been proved to be a feasible method for the synthesis of high purity ultrafine particles with controlled morphology and uniform distribution [17]. In the present work, ferroelectric $BaTi_{0.8}Sn_{0.2}O_3$ (BTSO) and ferromagnetic $La_{0.5}Ca_{0.5}MnO_3$ (LCMO) nanocristalline presenting both a curie temperature near room temperature (Tc=-24°C) were chosen and synthesized by the sol-gel techniques. To study the impact of different compositions of LCMO in the BTSO, structural, microstructural, dielectric and magnetic measurements on the multiferroic 0-3 and 3-0 type composite system (1-x) BTSO-(x) LCMO with (x =0, 0.3, 0.5, 0.7, 0.9 and 1) were performed for the first time.

## 2. Experimental

Polycrystalline (1-x) $BaTi_{0.8}Sn_{0.2}O_3$ - (x) $La_{0.5}Ca_{0.5}MnO_3$ composites with x = 0, 0.3, 0.5, 0.7, 0.9 and 1 were prepared in two stages. Firstly, the manganite $La_{0.5}Ca_{0.5}MnO_3$ was synthesized using the sol-gel technique. Stoichiometric amounts of analytical grade nitrates $La(NO_3)_3.6H_2O$, $Ca(NO_3)_2.4H_2O$ and $Mn(NO_3)_2.4H_2O$ were used as starting materials, while citric ($C_6H_8O_7$) and tartaric ($C_4H_6O_6$) acids were used as chelating agents. The mixture was dissolved in distilled water under vigorous stirring at room temperature in a beaker until a colorless and transparent solution was obtained. After that, the solution was transferred into a boiling flask and heated at 90°C using an oil bath for 7 hours. By removing the solvent, a homogeneous transparent sol was formed and a yellowish golden gel was obtained. The as prepared powder was heat-treated at 850°C for 8h/2°C min$^{-1}$ in air. Secondly, the Sn-doped BaTiO$_3$ was synthesized using the sol gel technique, barium acetate (Ba(CH$_3$COO)$_2$), and tin chloride (SnCl$_4$) were dissolved in acetic acid and titanium isopropoxide (C$_{12}$H$_{28}$O$_4$Ti) was added. After that a milky solution was obtained to which the ammonium hydroxide (NH$_4$OH) was added while heating at 80°C for about 1h until getting a transparent solution. A gel is obtained after evaporating the excess of the solvent, and the resulting powder was calcined at 1000 °C for 8h. The composites are fabricated by mixing BTSO and LCMO calcined powders in various desired wt% ratios of BTSO: LCMO,100:0, 70:30, 50:50, 30:70, 10:90 and 0:100 and then pressed into cylindrical pellets without using any binder and sintered at 1300°C for 2h. Hereafter, these composites are referred to as BL 0, BL 30, BL 50, BL 70, BL 90 and BL 100 for x= 0, 0.3, 0.5, 0.7, 0.9 and 1, respectively. The formation of the desired composites was verified by X-ray diffraction using an X'Pert PRO diffractometer (Cu-Kα, wavelength λ=1.5405 Å) operated at 40 kV and 40 mA. The cell parameters were refined from the diffraction patterns using the Fullprof software. Surface morphology and EDX spectra were observed by using scanning electron microscope (Tescan Vega 3 SEM) and by using a high resolution JEOL Field Emission Gun-Scanning Electron Microscope (FEG-SEM). The measurements of the magnetization vs. magnetic field M(H) were carried out by using a Physical Property Measurement System (PPMS- DynaCool) Quantum Design apparatus operating in temperature range 2-300 K and Magnetic field in the range 0_9 T. Magnetization was measured by using VSM (vibrating sample magnetometer) method that is integrated in this system. Dielectric properties were determined by using Hioki Impedance Analyzer.

## 3. Results and discussion

### 3.1. Chemical analysis

The percentage of the Mn$^{3+}$ and Mn$^{4+}$ ions in the magnetic material are checked quantitatively by using the oxidation-reduction method. For this purpose, LCMO powder was dissolved in dilute sulfuric H$_2$SO$_4$ and oxalic dehydrate H$_2$C$_2$O$_4$ acids under moderate heating. By reaction with the oxalate C$_2$O$_4^{2-}$ ions, both Mn$^{3+}$ and Mn$^{4+}$ were reduced to Mn$^{2+}$, and the excess of oxalic acid is then back titrated by using a 0.2 mol.dm$^{-3}$ KMnO$_4$ solution, the experimental results are found to be 48.67% for Mn$^{3+}$ and 51.33 for Mn$^{4+}$. These results agree with theoretical data, the oxygen stoichiometry was analyzed as described by Yang et *al.* [18], and it was found to be equal to 3.0066, confirming the stoichiometry of our sample.

### 3.2. X-ray diffraction

Fig. 1 shows the structural refinement of the prepared composite samples (1-x) BTSO - (x) LCMO with (x = 0, 0,3, 0,5, 0,7, 0,9 and 1) performed using the software FULLPROF [19]. No impurity phases were detected, and the results revealed the coexistence of individual phases, (cubic) BTSO and (orthorhombic) LCMO crystalline structures, eliminating the suggestion of the evolution of any new phase when sintering the composites. The XRD refinements of BTSO and LCMO for (x=0, 0.3, 0.5, 0.7, 0.9, 1) systems are presented in Table 1 and the obtained results are in good agreement with the wt% ratios used in the synthesis of composites. The structural estimated parameters by Rietveld refinements are collected in Table 2. At the same time, the XRD peak intensity ratios of BTSO to LCMO diffraction peaks are calculated and found to be equal to 1.7048, 1.5046, 1.2572 and 0.7152 for the compositions with x=0.3, 0.5, 0.7 and 0.9 respectively, showing the linear evolution of the peak intensity ratio as a function of x (Table 1). Besides, the characteristic peaks of BTSO and LCMO phases indicated in the XRD patterns by asterisks (*) and sharps (#) remain relatively independent of the composite materials revealing the negligible reaction between the phases. These results confirm the successful preparation of the 0-3 type connectivity for BTSO/LCMO multiferroic composite material.

Table 1. The percentage fractions of composites found by the Rietveld refinement, the intensities ratios and the average grain size.

|  | Fraction % | | | | | |
| --- | --- | --- | --- | --- | --- | --- |
|  | x=0 | x=0.3 | x=0.5 | x=0.7 | x=0.9 | x=1 |
| LCMO | 0 | 31.92 | 50.80 | 71.87 | 92.89 | 100 |
| BTSO | 100 | 68.08 | 49.20 | 28.13 | 07.11 | 0 |
| $I_{(BTSO)}/I_{(LCMO)}$ | - | 1.7048 | 1.5046 | 1.2572 | 0.7152 | - |
| grain size | 0.54µm | 1.07µm | 0.76µm | 0.69µm | 0.65µm | 66nm |

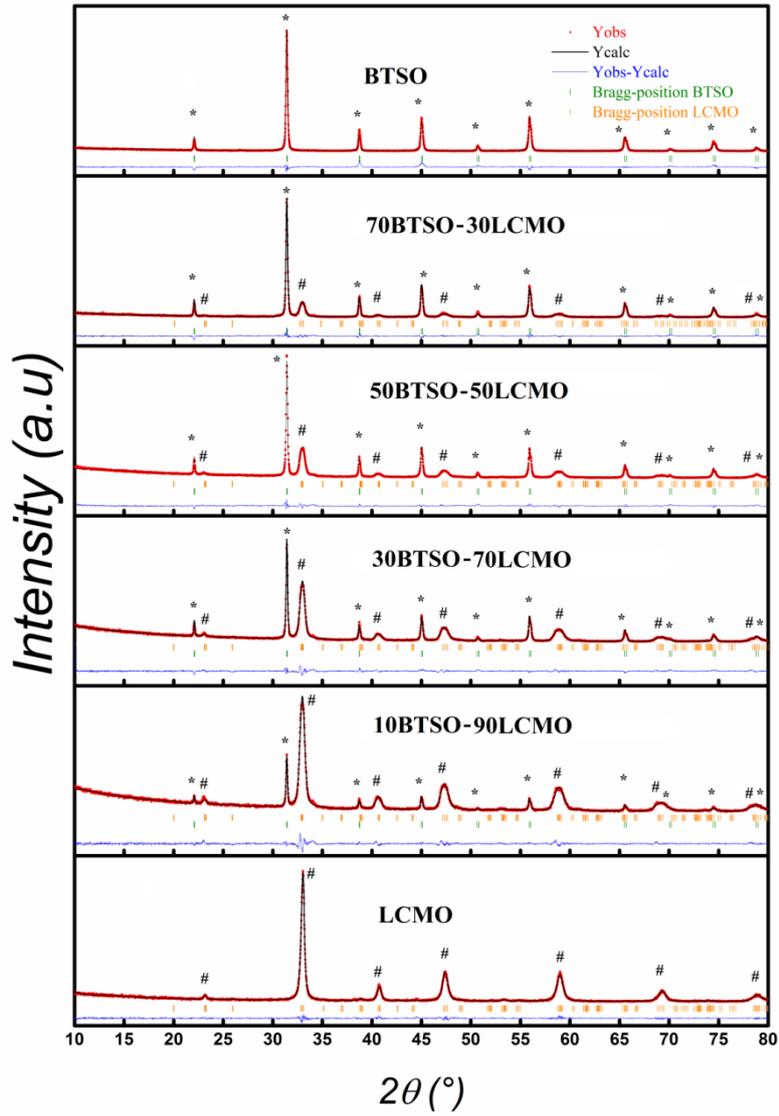

Fig.1. XRD patterns of the (1-x) BTSO – (x) LCMO composites with (x=0, 0.3, 0.5, 0.7, 0.9 and 1).

Table 2. Rietveld refined XRD parameters for bare BTSO, LCMO and BTSO-LCMO composite samples at room temperature.

| Sample | Space group | Lattice parameters Å | cell volume Å³ | $\chi^2$ | $R_P$ | $R_{wP}$ |
|---|---|---|---|---|---|---|
| BTSO | Pm3m | a= 4.0260 | V=65.2552 | 2.08 | 22 | 24.5 |
| 70BTSO-30LCMO | Pm3m<br>Pnma | a=4.0260<br>a=5.4671<br>b=7.6435<br>c=5.4243 | V=65.2582<br>V=226.6700 | 1.43 | 18.5 | 21.4 |
| 50BTSO-50LCMO | Pm3m<br>Pnma | a=4.0262<br>a=5.4375<br>b=7.6519<br>c=5.4484 | V=65.2675<br>V=226.6915 | 2.47 | 17.8 | 16.9 |
| 30BTSO-70LCMO | Pm3m | a=4.0263 | V=65.2715 | 3.51 | 22.3 | 22.0 |
| | Pnma | a=5.4271<br>b=7.6479<br>c=5.4661 | V=226.8181 | | | |
| 10BTSO -90LCMO | Pm3m | a= 4.0269 | V=65.3012 | 1.56 | 19.3 | 15.9 |
| | Pnma | a= 5.4404<br>b= 7.6514<br>c=5.4488 | V=226.8184 | | | |
| LCMO | Pnma | a= 5.4215<br>b=7.6493<br>c=5.4539 | V=226.1735 | 0.91 | 15.6 | 16.7 |

### 3.3. Microstructural study

The SEM images, presented in Fig. 2 (a)-(f), show that pure materials BL100 and BL0 exhibit spherical shaped grains with an average grain size of approximately 66 nm and 0.54µm, respectively, whereas in the composites, it is observed to be equal to 1.07 µm, 0.76 µm, 0.69 µm and 0.65 µm for BL30, BL50, BL70 and BL90 respectively (Table 1). An agglomeration of the particles is observed in Fig.2 (d and e) while an aggregation of the agglomerated particles is found in Fig.2 (b and c), showing grains with larger grain sizes which can affect the dielectric and ferroelectric properties of the ceramics. Indeed, an increase in the average grain size of the composite can generate a decrease in the specific surface area and in the proportion of grain boundaries, which could lead to a decrease in the contact surface between the ferroelectric and ferromagnetic grain materials limiting the appearance of interfacial polarization.

Furthermore, it is reported that large grain ceramics often show better ferroelectric characteristics compared to fine grain ceramics [20]. The reason is that large grain ceramics provide improved domain mobility and polarization, resulting in a higher degree of dipole movement and therefore, better electrical properties [21,22]. This can also affect the ferroelectric properties of the composites because of BTSO material shows hysteresis loop, whereas, the composite materials does not show any ferroelectric response which could be due to the increase in conductivity of the composites with increasing LCMO rate limiting the ferroelectric properties. The composition of the composites is confirmed by the appearance of La, Ca, Mn, Ba, Ti, Sn, and O peaks in the energy-dispersive X-ray spectroscopy (EDS) pattern (Fig. 3). Moreover, homogeneous dispersion of these elements on the surface of the

BL 70 composite is clearly shown in the elemental mapping plotted in Fig. 3 revealing good compatibility between BTSO and LCMO powders

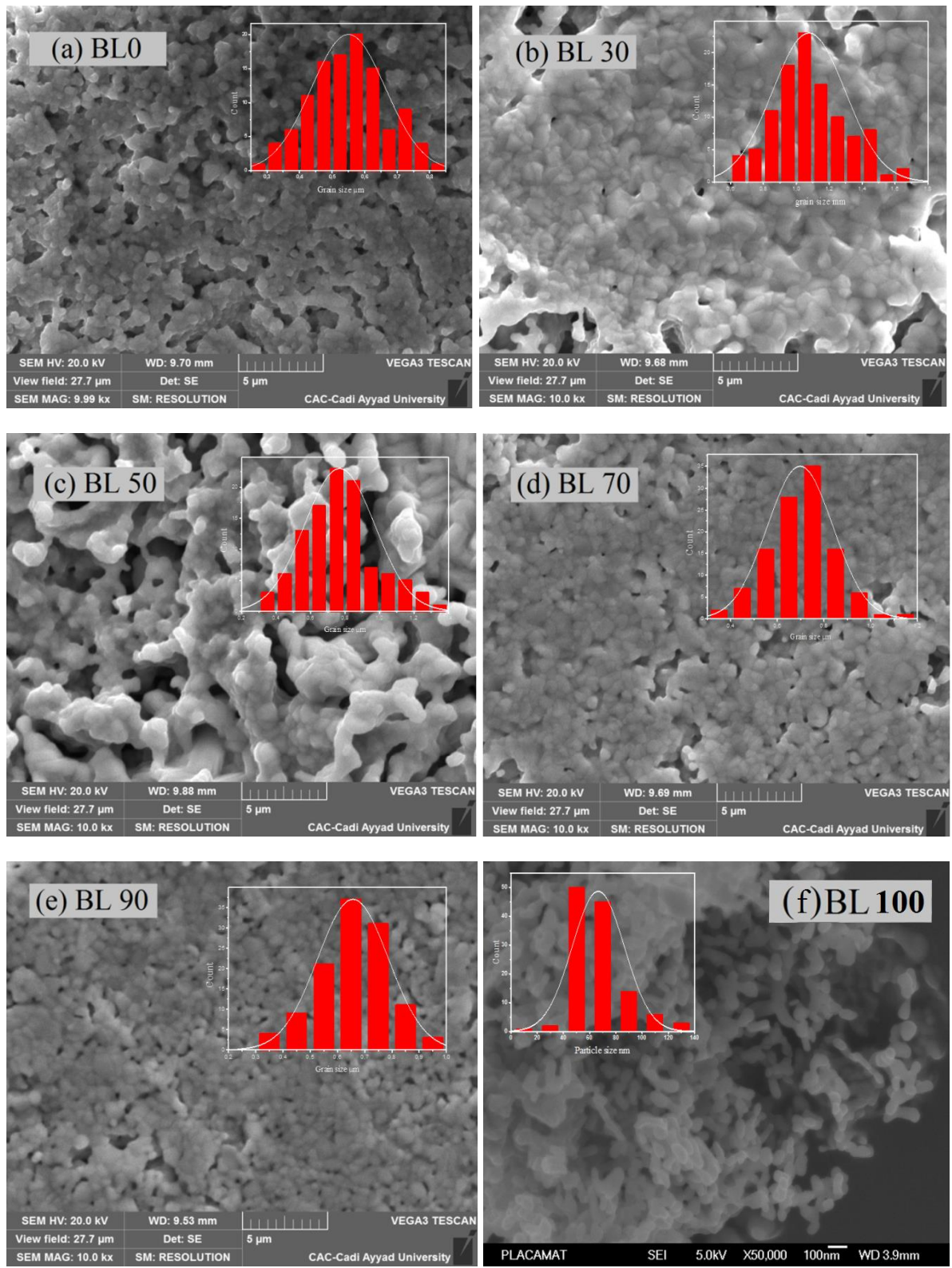

Fig. 2. The SEM images of the (1-x) BTSO-(x) LCMO composite ceramics (a) x=0, (b) x=0.3, (c) x=0.5, (d) x=0.7, (e) x=0.9, (f) x=1.

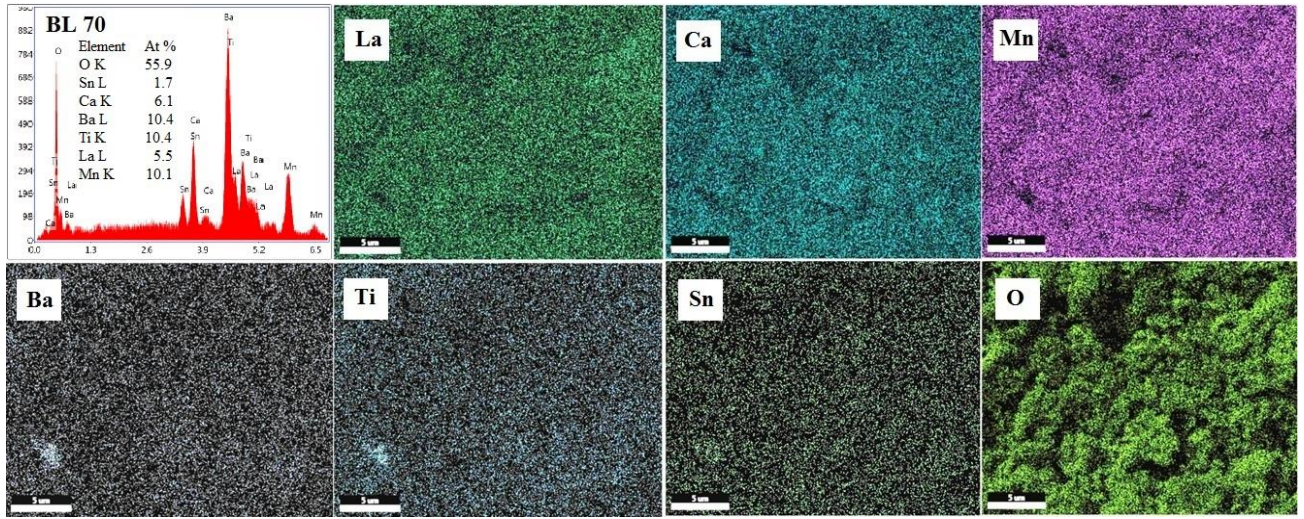

Fig. 3. Elemental mapping of Lanthanum (La), Calcium (Ca), Manganese (Mn), Barium (Ba), Titanium (Ti), Tin (Sn) and oxygen (O) on the surface of BL70 composite.

### 3.4. Dielectric properties

To study the effect of increased LCMO content on the dielectric properties of composites, measurements were made on pellet samples. Fig. 5 (a)-(e) shows the temperature dependencies of the dielectric constant and loss tangent for the composite samples at specific representative frequencies. A dielectric peak in the temperature vs dielectric constant curve can be observed at approximately Tc=-24°C in the parent compound BL0 (x=0), highlighting the ferroelectric to paraelectric phase transition, unlike composite materials show a simple anomaly at higher temperatures at about -9, -6 and -6°C for samples with x = 0.3, 0.5 and 0.7 respectively. It should be noted that the dielectric constant of the composite samples at the ferroelectric-paraelectric transition temperature for a representative frequency of 10 kHz was significantly increased and was observed to be 1660, 2262, 7700 and 11150 for samples with x = 0, 0.3, 0.5 and 0.7 respectively. The higher value of the dielectric constant may be attributed to the presence of optimal heterogeneity in composites [23] translated by the presence of a boundary layer between the ferroelectric (BTSO) and the ferromagnetic (LCMO) phases that give rise to the space charge interfacial polarization contributing to higher values of the dielectric constant. Besides, we note that the dielectric constant decreases at high frequency accompanied by small dispersion at the level of the peak anomaly, which can be explained by the Maxwell-Wagner interfacial polarization model which plays a key role in defining such systems [24,25]. Comparing to the other samples, we notice that the BL70 and BL90 show different dielectric temperature dependence where the dielectric permittivity reaches negative values for all the measuring frequencies at around 100°C and -

30°C for the samples respectively. It is axiomatic that the applied external electric field causes an oscillating electronic movement in the ceramic, if the electron oscillations are consistent with the direction of the applied external electric field, in this case, the direction of the polarization vector is consistent with the direction of the applied electric field and the value of the dielectric constant is positive. Accordingly, the negative value of the dielectric constant can be explained by oscillations of electrons which are opposite to the direction of the applied external electric field, i.e. the direction of polarization is opposite to the direction of the applied electric field. Consequently, the dielectric constant can be negative indicating a change in the electrical character of the material, i.e., the electrical nature of the material changes from capacitive to inductive [26]. The same behavior and explanation was reported by J.A. Bratkowska et *al* in Aurivillius-type $Bi_6Fe_{2-x}Mn_xTi_3O_{18}$ ceramics underlining the enhancement of the phenomenon with increasing the manganese doping content [27] and in multiferroic $BiFeO_3$–$PbFe_{1/2}Nb_{1/2}O_3$ ceramics [28]. Several explanations have been also reported to explain this phenomenon, such as the attribution of this dielectric behavior to thermal excitation of electrons to shallow trap levels inside the band gap [29] or also this can be related to the specific depolarization effects due to the motion of the domain walls in which the origin of the negative capacitance can be explained by the fact that in a dielectric material and in the absence of the applied electric field, the domains have different orientations up and down with equal sizes, therefore both the average polarization and the average depolarization field are zero. The application of an electric field displaces the wall domains unbalancing the sizes of the domains, this cause on the onset of the average polarization, oriented along the applied field. In its turn this generates a depolarization field which is opposite to the polarization. Consequently, the total field is counter-directed to the polarization which is a characteristic of the negative permittivity [30,31]. The dielectric loss tangents (tan δ) are also severely influenced by the conductivity of the materials. As the LCMO concentration increases in the composites, the dielectric loss of the composites increases, this can be explained by the high conductivity of the LCMO ferromagnetic phase. According to the principle of energy conservation, there must be a mechanism that allows energy to be stored and restituted only under certain conditions of temperature and frequency. From where the charges stored in the composite can be the source of the negative dielectric loss, they can be accumulated inside the material in the boundaries or can be anchored via non-bonding orbitals [32,33]. As a result of these charges separation, energy can be stored and can be emitted lately when specific frequency and temperature conditions are verified.

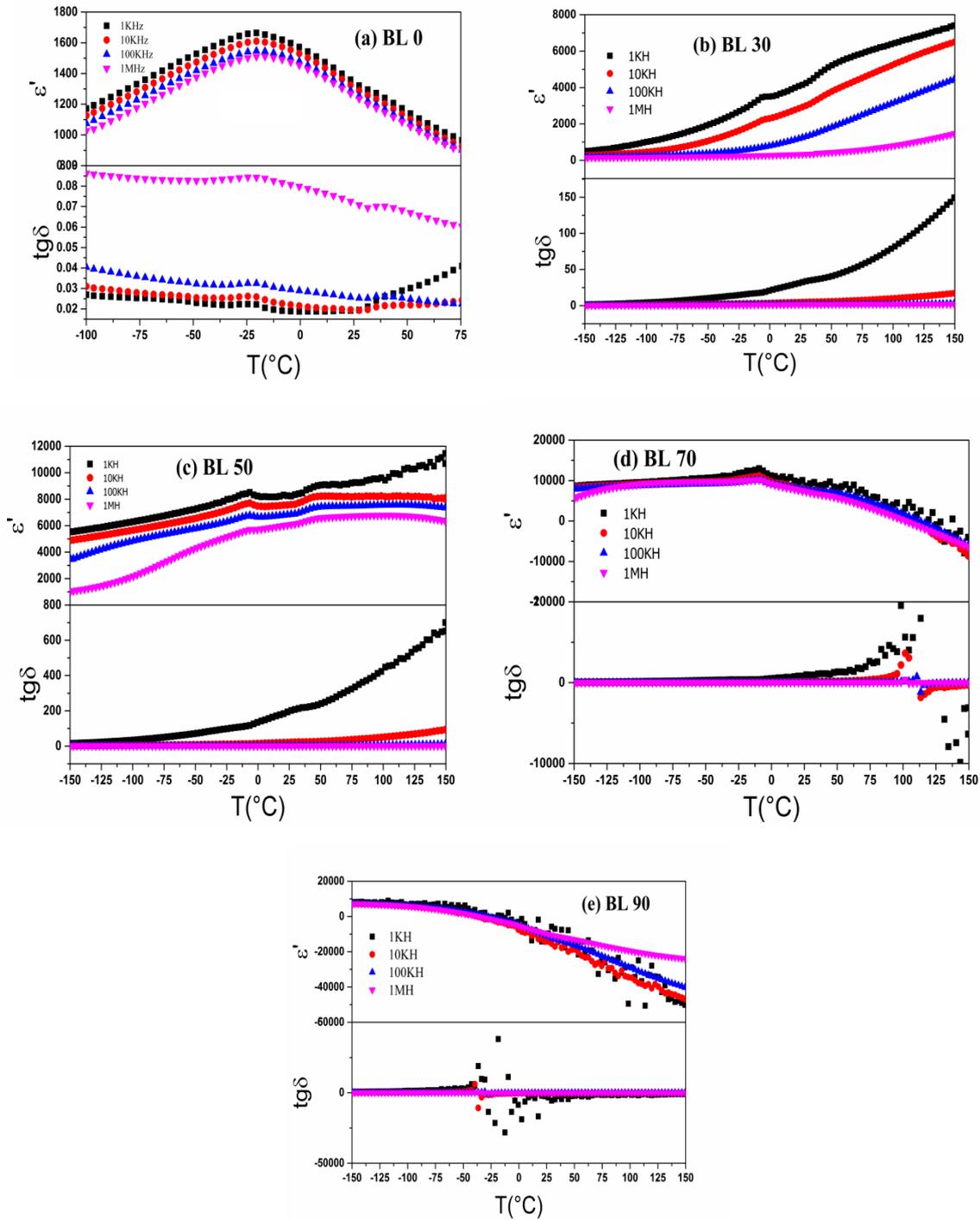

Fig. 5. Dielectric constant and dielectric loss as a function of temperature for the (1-x) BTSO-(x) LCMO samples: (a) x=0, (b) x=0.3, (c) x=0.5, (d) x=0.7 and (e) x=0.9.

3.5. Magnetic properties

The M-H loops of the composites were measured at the temperature of 2K. The results of measurements are presented in Fig. 6(a). The M-H loop of BTSO is not shown as it does not show a noteworthy magnetic behavior as compared to other compounds. The M(T) curve that was obtained at field 1KOe is plotted in Fig. 6(b), (c) and (d), where the ferromagnetic-paramagnetic phase transition is highlighted. The Curie temperature Tc was determined precisely by the magnetization derivative curve (dM/dT) obtaining 256 K, 121 K and 101 K for BL100, BL90 and BL70, respectively. As the BTSO material is non-magnetic, it is expected that the transition temperature Tc remains constant, whereas it decreases with increasing the BTSO content. This reduction is likely since the BTSO material is acting as a non-magnetic separation between the grains of LCMO underlining the strain induced by the grain boundary layer between BTSO and LCMO materials. According to our results, the presence of the non-magnetic BTSO material generates a magnetic anisotropy, which may be the reason for the decrease in the transition temperature. The same result was observed by E. Bose *et all* in LCMO/BTO composite materials [34]. All the composites exhibit typical ferromagnetic hysteresis loops and thus confirm the existence of an ordered magnetic structure in the composite samples. As expected, we notice that the obtained values of saturation magnetization (Ms) are lower than that of the pure LCMO (see Table 3). The lower value of saturation magnetization can be explained by the existence of the ferroelectric BTSO phase resulting in the dilution of magnetic properties [35]. Magnetoelectric coupling coefficient measurements for the BL30 sample are investigated using a dynamical method [36]. The sample was firstly poled at 60 K in the field of 10 kV/cm. Then the magnetoelectric polarization induced by a weak *ac* magnetic field of 0 - 0.64 Oe was measured as a function of the *dc* bias field at the temperatures 50K Fig. 7. At these low temperatures, the magnetoelectric response dominates by the quadratic paramagnetoelectric contribution (linear increase of the magnetoelectric current with magnetic field increase) according to magnetic measurements data for this sample, which shows mainly paramagnetic behavior with very small remanent magnetization (inset of Fig. 6 a). Estimated magnetoelectric coupling coefficient at H=10 kOe is only 1.3 ps/m. It is three times smaller than that in classical magnetoelectric $Cr_2O_3$. This relatively low magnetoelectric coupling in the BL30 composite is caused by the following factors: (i) small polarization and remanent magnetization; (ii) sizeable electric conductivity at T>100 K; (iii) large porosity of the ceramic composites. All these factors can be improved by tuning the synthesis process.

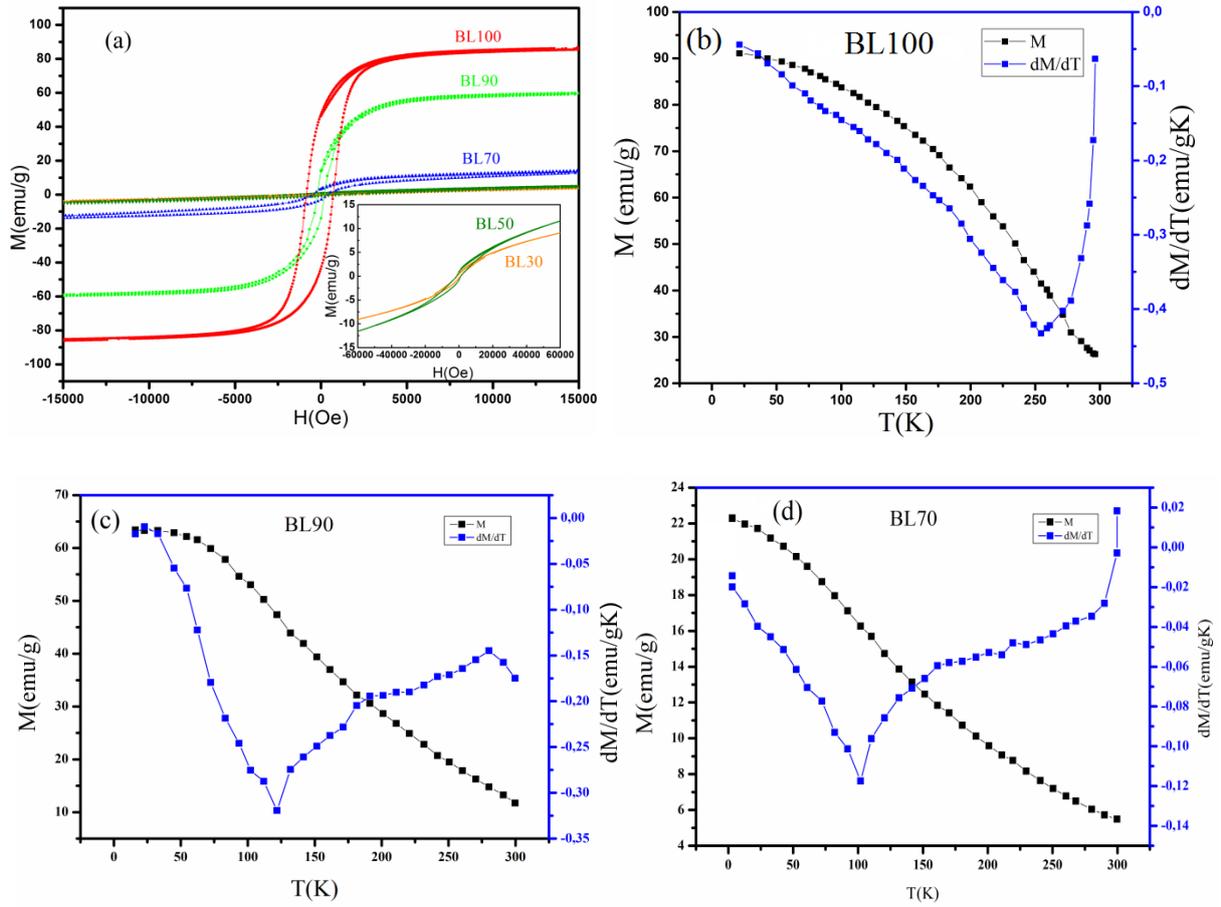

Fig. 6. (a) Magnetic hysteresis (M-H) loops of the BTSO-LCMO composites with different LCMO mass ratios (inset shows a zoom on both BL50 and BL30 samples), (b), (c) and (d) magnetization as function of temperature for pristine LCMO, BL70 and BL90.

Table 3. Remanent and saturation magnetization of the composite samples.

| Sample | Mr (emu/g) | Ms (emu/g) |
|---|---|---|
| 70% BTSO – 30% LCMO | 00.44 | 09.00 |
| 50% BTSO – 50% LCMO | 00.98 | 11.52 |
| 30% BTSO – 70% LCMO | 03.47 | 22.27 |
| 10% BTSO – 90% LCMO | 14.14 | 63.62 |
| 0% BTSO – 100% LCMO | 46.46 | 91.33 |

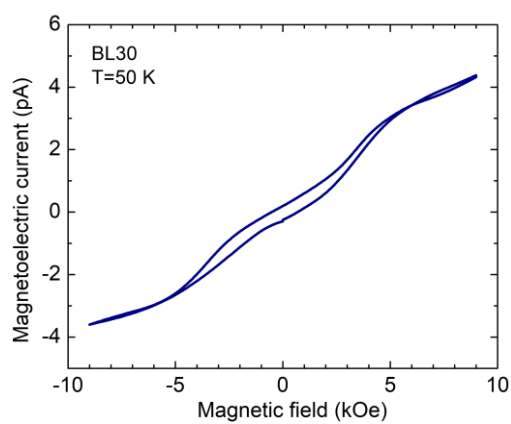

Fig. 7. Magnetoelectric current as function of the bias magnetic field in the BL30 composite ceramics

## 4. Conclusions

In summary, (1-x) BTSO -(x) LCMO (0≤x≤1) 0-3 and 3-0 type composite ceramics have been successfully synthesized by the sol-gel technique. From XRD and Rietveld refinement, we concluded that the composite samples of this series consist only of the two major BTSO and LCMO phases. Microstructural analysis reveals that the grain size of the composites is slightly decreased when increasing (x) ratio. The increase of LCMO content increases the dielectric constant, and at high (x) ratios, the electric character of the composite is changed from capacitive to inductive, exhibiting negative values of the dielectric constant as a result of the inverse polarization of the material. The magnetoelectric coupling coefficient measurements for the BL30 sample reveal weak response, which can be improved by tuning the synthesis process. All the composites exhibit ferromagnetic and ferroelectric character showing their multiferroic properties.

## Acknowledgements

The authors gratefully acknowledge the financial support of the European H2020-MSCA-RISE-2017-ENGIMA action, Slovenian research agency program P1-0125 and the CNRST Priority Program PPR 15/2015